\documentclass[aps,twocolumn,prl,showpacs,superscriptaddress]{revtex4}% Physical Review
\usepackage{graphicx}% Include figure files
\usepackage{dcolumn}% Align table columns on decimal point
\usepackage{bm}% bold math
\usepackage{amsmath}% More mathematical features
\usepackage{epsfig}
\usepackage{pstricks,pst-grad,pst-plot}% PsTricks

\begin{document}

%\preprint{LMS}

\title{Training-induced criticality in martensites}

\author{Francisco-Jos\'e P\'erez-Reche}
\affiliation{Laboratoire de M\'ecanique des Solides, CNRS UMR-7649, Ecole
Polytechnique, Route de Saclay, 91128 Palaiseau, France}
 \affiliation{Dipartimento di
Metodi e Modelli Matematici per le Scienze Applicate,\\  Universit\`a di Padova, Via
Trieste 63, 35121 Padova, Italy}
%
%\email{...}
%
\author{Lev Truskinovsky}
\affiliation{Laboratoire de M\'ecanique des Solides, CNRS UMR-7649, Ecole
Polytechnique, Route de Saclay, 91128 Palaiseau, France}
%
%\email{...}
%
\author{Giovanni Zanzotto}
\affiliation{Dipartimento di Metodi e Modelli Matematici per le Scienze Applicate,\\
Universit\`a di Padova, Via Trieste 63, 35121 Padova, Italy}

\begin{abstract}
  We propose an explanation for the self-organization towards
  criticality observed in martensites during the cyclic process known
  as `training'. The scale-free behavior originates from the interplay
  between the reversible phase transformation and the concurrent
  activity of lattice defects. The basis of the model is a continuous
  dynamical system on a rugged energy landscape, which in the
  quasi-static limit reduces to a sandpile automaton. We reproduce all
  the principal observations in thermally driven martensites,
  including power-law statistics, hysteresis shakedown, asymmetric
  signal shapes, and correlated disorder.
\end{abstract}


\maketitle

%\date{\today}

Experiments in martensites reveal intermittent behavior with power-law
statistics
\cite{Vives1994a,Carrillo1998,PerezReche2004Cyc,PerezReche2004NMG},
showing an intrinsic complexity comparable to that of turbulence,
earthquakes, internet networks, and financial markets. Criticality is
known to be an issue of great significance in contemporary science,
giving a framework for understanding the emergence of complexity in a
variety of natural systems \cite{Bak1987,Sornette2000}.  Within
materials science, criticality has been recognized in the last years
as a key factor in crystal plasticity, brittle fracture and damage
\cite{Alava2006Fracture-Zaiser2006Plasticity}. In the present paper we
develop a model explaining why similar behavior is also observed in
martensitic transformations.

Reversible martensitic transformations involve a coordinated
distortion of the crystal lattice and belong to the class of
ferroelastic first-order phase changes with athermal character
\cite{Olson1992,PerezRechePRL2005}. In such systems the macroscopic
strain discontinuity typically splits into a set of bursts
(avalanches) corresponding to transitions between neighboring
metastable states. The individual avalanches can be detected through
the measurement of the intermittent acoustic or calorimetric
signals. The size distribution observed in shape-memory alloys
(Cu-Al-Ni, Cu-Zn-Al, Cu-Al-Mn, Ni-Mn-Ga) was shown to be scale free
\cite{Vives1994a,Carrillo1998,PerezReche2004Cyc,PerezReche2004NMG,PerezRechePRL2005}.
Despite the apparent similarity with driven ferromagnetic systems,
where the scale-free Barkhausen noise has been known for a long time,
the experiments on memory alloys show features not observed in
magnets, and which are instead reminiscent of plastic shakedown. In
particular, the critical character of the avalanches
\cite{Carrillo1998,PerezReche2004Cyc} and the smoothing of the
hysteresis profile \cite{Miyazaki1999,PerezReche2004Cyc} emerge only
after multiple thermal cycling through the transition.

The mechanism leading to training-induced critical behavior in
martensites strongly resembles the phenomenology associated with
self-organized criticality (SOC) \cite{Bak1987}. The SOC paradigm in
the form of a sandpile automaton has been applied to martensitic
transformations in \onlinecite{Goicoechea1994}; which however, lacked
a connection to the physics of martensitic transformations. A
different set of models exploited the similarity between martensites
and magnetics by interpreting both in an Ising-type framework, with
zero temperature and quenched disorder \cite{Sethna_review2004}. In
this context the power-law over a few decades of avalanche sizes is
viewed as a sign of proximity of the system to a classical critical
point. Criticality then emerges only as a result of the external
tuning of the disorder. Furthermore, the symmetric avalanche shapes
with scaling collapse which are expected in these models
\cite{Sethna_review2004}, contrast those experimentally recorded
\cite{PerezReche_TechRep2004}. More recent modeling has focused on the
direct simulation of martensitic transformations within the framework
of elasticity theory \cite{Sreekala2004}. While the corresponding
numerical tests show some scaling in avalanche sizes, the system is
unable to memorize its state of disorder upon unloading, failing
therefore to exhibit the effects of training.

A key experimental observation left aside in the preceding theoretical
work is the dislocational activity assisting the development of the
phase transformation in these materials
\cite{RiosJara1987I,Pons1990,Lovey1999,Miyazaki1999,Cuniberti_DSC2004}. It
has been repeatedly observed that dislocations are indeed introduced
in shape memory alloys during training. In particular, under periodic
driving the degree of defectiveness first increases monotonically and
then saturates \cite{Pons1990,Lovey1999}. Our model shows that this
dislocational activity is highly correlated and is ultimately
responsible for the scale-free character of the reversible behavior of
martensites. More precisely, the attainment of criticality is due to
the ability of the crystal to develop an optimal amount of disorder.

%%%%%%%%%%%%%%%%%%%%%%%%%%%%%%%%%%
\begin{figure}
  \centering \epsfig{file=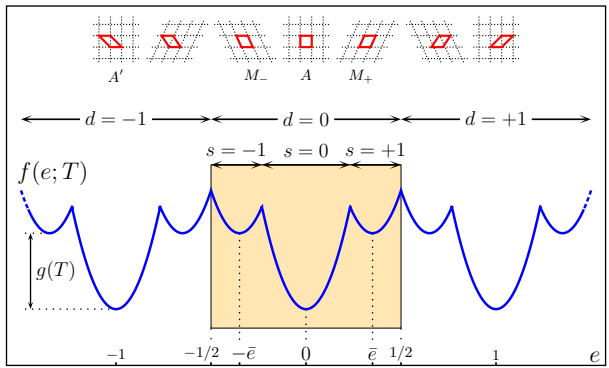,width=8cm,clip=} 
\caption{ \label{fig:energy} Schematic
    representation of the periodic potential $f(e;T)$ with the shaded
    box highlighting a single domain of periodicity. Insets on top
    show the lattice structures corresponding to the bottoms of the
    potential wells, with lattice cells marked for the austenite $A$
    and for the martensite variants $M_{\pm}$. The austenite $A'$
    corresponds to a different period of the potential.}
\end{figure}
%%%%%%%%%%%%%%%%%%%%%%%%%%%%%%%%%%%
To describe in an unified way the processes involving both
dislocations and phase boundaries, we consider a prototypical 2D
system of kinematically compatible elastic units resulting from a
suitable triangulation of a square lattice (see an example of such
procedure in \onlinecite{Conti2004}). To each unit with index $i$ we
assign a multi-well strain energy function depending on a single
scalar order parameter $e$ which, in turn, is a combination of the
components of the discrete strain tensor. The adiabatic elimination of
the harmonic non-order-parameter strain variables by means of the
equilibrium equations and the kinematic compatibility constraints
leads to a non-local elastic energy of the form
\begin{equation}
\label{total_discretized_energy} \widetilde{\Phi}(e;T)=\sum_i f(e_i;T)+\frac{1}{2}
\sum_{i,j}K_{ij}e_i e_j,
\end{equation}
where ${\bf K}=\{K_{ij}\}$ is the kernel of the long-range elastic
interactions, and $f$ is a periodic function as in
Fig.~\ref{fig:energy}. In each period we use the three-parabolic
approximation $f(e;T) =\frac{1}{2}(e-w)^2+ g(T) s^2$, where $s=0$ in
the high symmetry phase (austenite) and $s= \pm1$ in the two variants
of the low symmetry phase (martensite). The parameter $w = d +
\bar{e}s$, where $\bar{e}$ is the transformation strain, defines the
location of the bottoms of the energy wells; the integer-valued
parameter $d$ specifies the period of $f$. We emphasize that no
randomness have been assumed in the model.

\begin{figure}[fh]
  \centering \epsfig{file=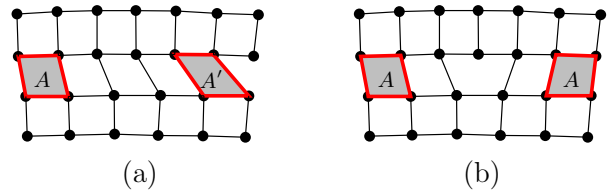,
    width=8cm,clip=} \caption{ \label{fig:dislocation} Two
    representations of a dislocation. (a) Continuous lattice
    deformation involving partial slip, as is interpreted in the
    text. (b) The same atomic configuration viewed within the
    classical interpretation involving a discontinuous deformation and
    a non-zero Burgers vector. The shaded units refer to austenite in
    the wells $A$ and $A'$ in Fig.~\ref{fig:energy}.}
\end{figure}

The global periodicity of the energy takes large shearing distortions
into account \cite{Ericksen1980-PitteriZanzotto2003}, so that both the
phase change and dislocation formation can be handled
simultaneously. When the transformation strain is small and the energy
barriers for slip are much higher than the barriers for the phase
transition and twinning (`weak transitions'), no lattice-invariant
shears occur (as with the R-phase of NiTi \cite{Miyazaki1999}, or the
pre-martensitic transformation in Ni-Mn-Ga \cite{PerezReche2004NMG})
and modeling can proceed according to the Landau theory. Such phase
changes do not generate significant dislocational activity and are
largely reversible, as is assumed, for instance, in
\cite{Sreekala2004,Shenoy06}. On the contrary, in `reconstructive
transformations' the transformation strain is large, as it lays at the
boundary of the periodicity domain (e.g. the ideal Bain transformation
from bcc to fcc). In those cases the energy barriers to slip are only
as high as the transformation barriers, and the phase change advances
`half way' towards the formation of dislocations \cite{Conti2004}. As
a result, defects proliferate making the transition irreversible
\cite{Bhattacharya2004}. In between these two extremes a range of
possibilities exists, where defect formation plays an increasing role
as the transformation strain gets closer to the boundary of the
maximal periodicity domain. In particular, all the martensitic
transformations considered in the experiments that we are concerned
with
\cite{Vives1994a,Carrillo1998,PerezReche2004Cyc,PerezReche2004NMG,PerezRechePRL2005},
involve a transformation strain $\bar{e}$ which is very close to the
ideal Bain strain \cite{Balandraud_Zanzotto2006}. When slowly driven,
these systems are expected to exhibit cell deformations not confined
to one periodicity domain, but rather extending to an unbounded
portion of the periodic energy landscape. This involves formation of
dislocations, which our kinematic compatibility assumption does not
exclude, as is exemplified in Fig.~\ref{fig:dislocation}, where we
compare the standard representation of a dislocation (b) with the one
adopted in this paper (a).
%
%%%%%%%%%%%%%%%%%%%%
%%%%%%%%%%%%%%%%%%%%

We drive the system quasi-statically trough the function $g(T)$, by
changing the temperature $T$. Since the transformations are typically
athermal \cite{PerezRechePRL2005}, we consider a purely mechanical
setting with overdamped dynamics. By making this assumption we treat
the acoustic emission as a part of dissipation.  In the limit of
infinitely slow driving, the dynamics projects on the local minima of
the total energy $\widetilde{\Phi}$, which form a discrete set of
branches $e=e(g)$, with $e^- < e < e^+$, where the extremes $e^{\pm}$
correspond to marginally stable configurations \cite{PT05}. For
piece-wise parabolic $f$ as in Fig.~\ref{fig:energy}, the limits
$e^{\pm}$ of each branch can be explicitly written as a function of
$g$ and $\bar{e}$. When such limits are reached, the instability
resolves through a fast event (avalanche) which brings the system to
another equilibrium branch. In this way the dynamics becomes
piece-wise continuous (see \onlinecite{TV04} for a study of the 1D
case).

We proceed by eliminating through minimization the linear elastic
strain $e$ at given $d$ and $s$ obtaining ${\bf e} = ({\bf 1} + {\bf
  K})^{-1}{\bf w}$. The relaxed energy is of the Ising type
\begin{equation}
\label{Hamiltonian_S_D} {\widehat{\Phi}}= -\sum_{i,j} \left[\frac{\bar{e}^2}{2}
(J_{ij}-\frac{2g}{\bar{e}^2}\delta_{ij}) s_i s_j+\frac{1}{2} J_{ij} d_i
d_j+\bar{e}J_{ij} s_i d_j\right],
\end{equation}
where the two discrete spin variables describe the phase
transformation ($s$) and the plastic slip ($d$), and ${\bf J}= ({\bf
  1} + {\bf K})^{-1}-{\bf 1}$. Since the energy
(\ref{Hamiltonian_S_D}) is supposed to penalize the inhomogeneity of
the field $w_i$ induced by either phase boundaries or dislocations, we
assume that the corresponding term has the particular form
$\frac{1}{4}\sum J_{ij} (w_i-w_j)^2$ so that $J_{ii}=-\sum_{j \neq
  i}J_{ij}$. We furthermore assume the kernel ${\bf J}$ to be of the
ANNNI type, to account for the competing interactions driving both the
coarsening and the refinement of the microstructure
\cite{ren2000,Shenoy06}. Specifically, we consider $J_{ ij} = J_1> 0$
for nearest neighbors, and $J_{ij} = -J_2 < 0$ for next-to-nearest
neighbors; the ensuing diagonal-dominated structure of the matrix
${\bf J}$ does not prevent the original matrix ${\bf K}$ in
(\ref{total_discretized_energy}) from being dense. Suitable
inequalities ensure the metastability of the individual equilibrium
branches \cite{TV04}; in particular, the condition $J_1 \geq 2 J_2$ is
sufficient for our automaton to reach a steady state (we use $J_1$ =
0.062, $J_2$ = 0.03 in the simulations presented below).

Under the above hypotheses, the piece-wise continuous dynamics becomes
a sandpile automaton, whose main variable is the elastic strain
$\delta_i=e_i-w_i$ representing the `local height'. Once a cell
becomes unstable (the condition $e^{-}<e_i<e^{+}$ is violated),
$\delta_i$ is updated as
\begin{eqnarray}
\label{Delta_i}
\delta_i &\rightarrow & \delta_i -4(J_1-J_2) r\\
\nonumber \delta_{j} &\rightarrow & \delta_{j}+J_1 r,\;\;
\mbox{\small{$j$ nearest neighbors of $i$}},\\
\nonumber \delta_{k} & \rightarrow & \delta_k-J_2 r,\;\; \mbox{\small{$k$
next-to-nearest neighbors of $i$}},
\end{eqnarray}
where $r=\pm \bar{e}$ for phase transitions, and $r=\pm(1-2\bar{e})$
for slips. Since each update may make new sites unstable, the updates
continue at constant $g$ until the system is fully equilibrated. For
$J_2 = 0$, $\bar{e} = \frac{1}{3}$, and $g = 0$ the automaton
(\ref{Delta_i}) reduces to the standard BTW sandpile \cite{Bak1987}.
%%%%%%%%%%%%%%%%%%%%%%%%%%%%%%%%%%%%%%
\begin{figure}[h]
\centering \epsfig{file=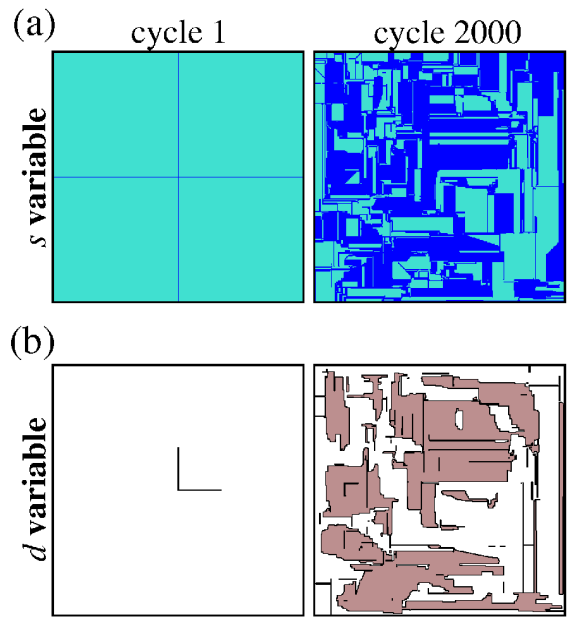, width=7.5cm,clip=} \caption{
  \label{fig:evolution_disorder} Evolution of the phase and defect
  microstructures in the lattice during thermal cycling. A minimal
  dislocation loop was placed at the middle of the system. (a)
  Configuration of the martensitic phase domains, represented by the
  field $s_i$, after cycle 1 and cycle 2000 (turquoise and blue
  indicate $s = 1$ and $s = -1$, respectively). (b) The corresponding
  configurations of the slip variable $d$. White and brown colors
  indicate $d=0$ and $d\neq0$ (mostly $|d|=1$), respectively. The
  black contours separate elements with different $d$.}
\end{figure}
%%%%%%%%%%%%%%%%%%%%%%%%%%%%%%%%%%%%%%%%

\begin{figure}
\centering \epsfig{file=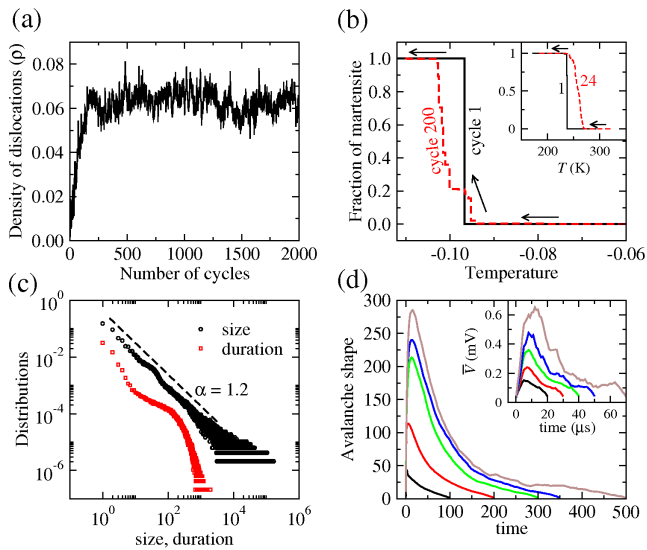, width=8.5cm,clip=}
\caption{\label{fig:other_features} Evolution to criticality and
  features of the critical state. (a) The dislocation density $\rho$
  during the first 2000 cycles. (b) Cooling curves representing phase
  fraction vs.~temperature in arbitrary units (cycle 1, solid line;
  cycle 200, dashed line). The corresponding experimental data
  \cite{PerezReche2004Cyc} are shown in the inset for comparison (1st
  and 24th cycles). (c) Steady-state power-law distribution of the
  avalanche sizes, and non-power-law distribution of durations
  (displaced one decade lower for clarity). (d) Asymmetric avalanche
  shapes. The inset shows the corresponding experimental data
  \cite{PerezReche_TechRep2004}.}
\end{figure}

We implement the model numerically on a $501\times501$ grid for an
almost reconstructive transformation with $\bar{e} = 0.47$ and open
boundary conditions. The initially homogeneous austenite is chosen to
contain only a minimal dislocation loop.  The crystal is then
thermally cycled through the complete transformation, $g$ being a
periodic triangular function of computational time. The intensity of
the acoustic bursts registered experimentally is linked to the size of
the avalanches (total number of updates before
stabilization). Fig.~\ref{fig:evolution_disorder}a shows the
development of the phase microstructure during the training
period. The level of plastic deformation is monitored through the
density $\rho$ of nearest neighbors with differing values of
$d_i$. Fig.~\ref{fig:evolution_disorder}b shows the formation of
dislocations induced by training and marked by the steep initial
increase in the variable $\rho$ (Fig.~\ref{fig:other_features}a). The
creation of correlated dislocation microstructure quickly saturates,
in accordance with the experiments \cite{Pons1990,Lovey1999}. In
Fig.~\ref{fig:other_features}b, we observe the smoothing effect of the
self-organized defects on the cooling curves (and hence on the
hysteresis cycle). The dislocational activity leads to the increase of
the martensite starting temperature, similarly to what is reported
experimentally \cite{PerezReche2004Cyc}. The parallel development of
criticality is indicated by the emergence of the power-law statistics
for the avalanche sizes (Fig.~\ref{fig:other_features}c). At the
beginning of the training period the avalanche distribution is
supercritical with a peak at large sizes evident from the sharp
initial cooling curves in Fig.~\ref{fig:other_features}b. The peak
eventually vanishes (Fig.~\ref{fig:other_features}c), as in the
experiments \cite{PerezReche2004Cyc}. Two further predictions of the
model matching experimental data \cite{PerezReche_TechRep2004} concern
the strong asymmetry of the avalanche shapes, and the absence of their
scaling collapse (see Fig.~\ref{fig:other_features}d). Similar effects
are also observed in Barkhausen noise, earthquakes, and dynamic
fracture (see the discussion in
\onlinecite{Zapperi_nature_physics2005}). Fig.~\ref{fig:other_features}c,
shows the distribution of avalanche durations (number of simultaneous
updates in an avalanche) predicted by the model, which deviates from a
power-law. This indicates that in the present framework a scale-free
size distribution does not always imply scaling in time. While this
prediction for an idealized system (with neither inertia nor fixed
pinning sites) disagrees with the scaling for durations reported
experimentally \cite{Vives1994a}, our model does generate an almost
power-law distribution of durations after the introduction of a small
amount of quenched disorder represented by a Gaussian distribution of
$\delta_i$ with zero average in the initial configuration. Such
modification, however, does not influence the power-law distribution
of avalanches, nor does it affect the pulse asymmetry.
%%%%%%%%%%%%%%%%%%%%%%%%%%%%%%%%%%%%%%%%%%%%%%%%%%%
%%%%%%%%%%%%%%%%%%%%%%%%%%%%%%%%%%%%%%%%%%%%%%%%%%%

When in the original setting with two variables $s$ and $d$ the phase
transformation is suppressed (no variable $s$), the model describes
the micromechanics of stress-driven intermittent plastic flow in
crystals \cite{Dimiduk2006}. In this case the system is defined by the
single integer-valued order parameter $d$ as in the phase-field
description of plasticity \cite{Koslowski2002}, to which our model
offers an analytically accessible automaton alternative
\footnote{Models with one variable and random thresholds have been
  explored in the literature (e.g.~\cite{Bak1987}). In our model such
  randomness is not postulated \emph{a priori} but is instead
  developing due to the complex coevolution of the two variables $s$
  and $d$. A two-variable model with well-separated relaxation times
  has been proposed in \cite{Coolen1993}.}. What is more, the present
scheme of plasticity does not require ad-hoc procedures for the
nucleation and annihilation of dislocations, as in discrete
dislocation dynamics \cite{MCMiguel2001}. When, in contrast, the
variable $d$ can be neglected in the original setting, as in weak
martensitic transformations (or in magnetics), the model still
generates several decades of power-law avalanche distribution in
accordance with the experiments in Ni-Mn-Ga \cite{PerezReche2004NMG},
but only in the presence of some quenched disorder represented again
by a Gaussian distribution of $\delta_i$ in the initial
configuration. In this case the emergence of limited scaling does not
involve training, and can be explained by the proximity of the system
to a classical critical point \cite{Sethna_review2004}.

In summary, the proposed model accounts for all the main observed
phenomena accompanying the training process in martensites leading to
criticality. The agreement with experiment clearly indicates that SOC
originates in these systems as a result of the interplay between the
reversible phase change and the irreversible development of an optimal
amount of plastic deformation.

We thank J. Aizenberg, P. Collet, Ll. Ma{\~n}osa, A. Planes, S. Roux,
G. Tarjus, and E. Vives for helpful comments. FJPR was supported by
the European contract MRTN-CT-2004-505226 and by Spanish grant MEC
EX2005-0792. G.Z. was partially supported by grants from PRIN2005 and
INdAM, Italy.

%%% BibTex bibliography %%%%%%%%%%
%\bibliographystyle{apsrev}
%\bibliography{../bibliography}
%%%%%%%%%%%%%%%%%%%%%%%%%%%%%%%%%%

\end{document}